\documentclass[12pt]{article}
\usepackage{amsfonts,amsmath}
\usepackage[hypertex]{hyperref}
\usepackage{mathrsfs}

\makeatletter \@addtoreset{equation}{section} \makeatother

\newcommand{\be}{\begin{equation}}
\newcommand{\ee}{\end{equation}}
\newcommand{\bee}{\begin{eqnarray}}
\newcommand{\beee}{\begin{array}}
\newcommand{\eee}{\end{eqnarray}}
\newcommand{\eeee}{\end{array}}

\newcommand{\ga}{\alpha}
\newcommand{\pa}{{\dot{\ga}}}
\newcommand{\pb}{{\dot{\gb}}}

\newcommand{\gb}{\beta}
\newcommand{\gga}{\gamma}

\newcommand{\G}{{\mathcal G}}

\newcommand{\W}{{\cal W}}

\newcommand{\Hh}{{\cal H}}

\newcommand{\rhs}{{\it r.h.s.} }

\newcommand{\ie}{{\it i.e.,} }
\newcommand{\ls}{\!\!\!\!\!\!}

\newcommand{\gvep}{\varepsilon}

\newcommand{\gs}{\sigma}

\newcommand{\go}{\omega}

\newcommand{\q}{\,,\qquad}

\newcommand{\nn}{\nonumber}

\newcommand{\half}{\frac{1}{2}}

\newcommand{\p}{\partial}
\newcommand{\D}{{\cal D}}

\newcommand{\f}{\frac}

\newcommand{\bu}{\bar{\kappa}}

\def\G{\Gamma}

\newcommand{\PPP}{ {J} }

\newcommand{\xx}{{\bf x}}

\def\Di{${\mathcal A}$}
\def\Ne{${\mathcal B}$}
\newcommand{\dgb}{{\dot \gb}}
\newcommand{\dga}{{\dot \ga}}

\newcommand{\dr}{{\rm d}}
\newcommand{\zz}{{\bf z}}

\newcommand{\HH}{\mathscr{H}}
\begin{document}

\begin{flushright}

{\small FIAN/TD/10-16}
\end{flushright}
\vspace{1.7 cm}

\begin{center}
{\large\bf Current Interactions and Holography from
the 0-Form Sector of\\
\vspace{0.1cm}
 Nonlinear Higher-Spin Equations
}

\vspace{1 cm}

{\bf  M.A.~Vasiliev}\\
\vspace{0.5 cm}
{\it
 I.E. Tamm Department of Theoretical Physics, Lebedev Physical Institute,\\
Leninsky prospect 53, 119991, Moscow, Russia}

\end{center}

\vspace{0.4 cm}

\begin{abstract}
\noindent The form of higher-spin current interactions in $AdS_4$ is
derived from the full nonlinear higher-spin equations in the sector
of Weyl 0-forms. The coupling constant in front of spin-one currents
built from scalars and spinors  as well as Yukawa coupling are
determined explicitly. Couplings of all other higher-spin current
interactions are determined implicitly. All couplings  are shown to
be independent of the phase parameter of the nonlinear higher-spin
theory. The proper holographic dependence of the vertex on the
higher-spin phase parameter  is shown to result from the boundary
conditions on the bulk fields.

\end{abstract}

\newpage
\tableofcontents

\newpage

\section{Introduction}
Higher-spin (HS) gauge theory is a theory of interacting massless
fields of all spins. Massless fields of spins $s\geq 1$ are gauge
fields supporting various types of gauge symmetries
 \cite{Frhs,Frfhs}. Gauge fields of spins $s=1$ and $s=2$ are familiar from the
conventional models of fundamental interactions, underlying the Standard Model and GR, respectively.
Symmetries of gauge fields of spins $s>2$ are HS symmetries. The
latter are anticipated to play a role at ultra high energies
possibly beyond  Planck energy. Since such energies are unreachable
by modern accelerators the conjecture that a fundamental theory
exhibits  HS symmetries at ultra high energies provides a unique
chance to explore properties of this regime. HS symmetries severely
restrict the structure of HS theory, underlying many of its unusual
properties.

One is that, as shown by A.~Bengtsson, I.~Bengtsson, Brink
\cite{Bengtsson:1983pd,Bengtsson:1983pg} and  Berends, Burgers, van
Dam \cite{Berends:1984wp,Berends:1984rq}, HS interactions consistent
with HS gauge symmetries contain higher derivatives (note however
that no higher derivatives appear at the quadratic level in a
maximally symmetric background geometry \cite{Frhs,Frfhs}). Another one,
originally indicated by the analysis of HS symmetries in
\cite{Berends:1984wp,Berends:1984rq} and later shown to follow from
the structure of HS symmetry algebras \cite{hsa4,OP1}, is that a HS
theory containing a propagating field of any spin $s>2$ should
necessarily contain an infinite tower of HS fields of infinitely
increasing spins. Since higher spins induce higher derivatives in
the interactions any HS theory with an infinite tower of HS fields
is somewhat nonlocal. Of course some kind of nonlocality beyond
Planck scale should be expected of a theory anticipated to capture
the quantum gravity regime.

Appearance of higher derivatives in interactions demands a
dimensionful  constant $\rho$ which was identified in
 \cite{Fradkin:1987ks,Fradkin:1986qy} with the radius of the background
$(A)dS$ space. In this setup, the higher-derivative
vertices admit no meaningful flat limit in agreement with numerous
no-go statements ruling out nontrivial interactions of massless HS
fields in Minkowski space \cite{Coleman:1967ad,Aragone:1979hx} (see
 \cite{Bekaert:2010hw} for more detail and references).

The feature that consistent HS interactions demand
non-zero cosmological constant
acquired a deep interpretation with the discovery of the $AdS/CFT$
correspondence \cite{Maldacena:1997re,Gubser:1998bc,Witten:1998qj}.
After general holographic aspects of  HS theory were pointed out
in \cite{Sundborg:2000wp,WJ,Sezgin:2002rt}, the precise
proposal on the $AdS_4/CFT_3$
correspondence was put forward by Klebanov and Polyakov
\cite{Klebanov:2002ja}. Its first explicit check
 was done by Giombi and Yin in \cite{Giombi:2009wh} triggering
sharp increase of interest in HS theories and HS holographic duality
\cite{Aharony:2011jz}-\cite{Gaberdiel:2010pz}.

Analysis of holographic duality in the HS theory is hoped to
shed light on the very origin of $AdS/CFT$ correspondence. Since HS gauge symmetry
principle is the defining property of HS theories, it is crucially
important to elaborate a manifestly gauge
invariant definition for the boundary generating functional.
A new proposal in this direction was recently put forward in
\cite{Vasiliev:2015mka} which, along with the idea of derivation
of holographic duality from the unfolded equations \cite{Vasiliev:2012vf}
respecting all necessary symmetries, may provide a useful setup for better
understanding of holographic duality.

Geometric origin of the dimensionful parameter $\rho$
has an important consequence that any HS gauge theory  with
unbroken HS symmetries does not allow a low-energy analysis because
a dimensionless derivative $\rho \f{\p}{\p x}$ that appears in the
expansion in powers of derivatives cannot be treated as small.
This is most obvious from the fact that the rescaled covariant derivatives
 $\D=\rho D$, which are non-commutative in the background $AdS$ space-time of curvature $\rho^{-2}$,
have commutator of order one, $[\D\,,\D] \sim 1$.

As a result, in a field redefinition
\be
\label{exp}
\phi \to  \phi' =\phi + \sum_{n,m=0}^\infty a_{nm} \D^n \phi \D^m \phi\,,
\ee
all terms with higher derivatives may give comparable contributions. So, whether
expansion (\ref{exp}) is local or not should depend solely on the behavior of the
coefficients $a_{nm}$ at $n,m\to \infty$. If at most a finite number of coefficients
$a_{n,m}$ is nonzero,  field redefinition (\ref{exp}) is genuinely local. Based on the results of this
paper, in  \cite{loc} we conjecture a restriction on the coefficients analogous to $a_{nm}$ appropriate
for the definition of the local frame in the twistor-like formalism of the
standard formulation of nonlinear HS equations of motion in $AdS_4$ \cite{more}.

Importance of the proper definition of locality was originally  raised in \cite{prok}
where it was shown that by a field redefinition of the form (\ref{exp}) it is possible
to get rid of the currents from the {\it r.h.s} of HS field equations. Recently
this issue was reconsidered  in
\cite{Vasiliev:2015wma,Boulanger:2015ova,Bekaert:2015tva,Skvortsov:2015lja}
(see also a subsequent paper \cite{Taronna:2016xrm}). In particular in
\cite{Vasiliev:2015wma} a proposal was put forward on the part of the problem
associated with the exponential factors resulting from so-called inner Klein
operators while the structure of the preexponential factors was only partially
determined. In \cite{loc}, the results of \cite{Vasiliev:2015wma} are
extended to the preexponential factors at the quadratic order.

Analysis of this paper was motivated by the work \cite{Gelfond:2010pm} where
current HS interactions were rewritten in the unfolded form. A natural
 question on the agenda was to derive these results from the nonlinear
HS equations of \cite{more}. This study was boosted by the recent
papers \cite{Boulanger:2015ova,Skvortsov:2015lja} where such an attempt  was undertaken.
Unfortunately, leaving the problem unsolved, the authors of
these papers arrived at some counterintuitive  conclusions  raising  the issue of
physical interpretation of the  HS equations of \cite{more}, thus
urging us to reconsider the problem.

  In this paper, confining ourselves to the 0-form sector of the HS equations,
 which is simpler than that of 1-forms, we present  a simple field redefinition
 that reduces the first nonlinear corrections resulting from the nonlinear HS  equations to
the usual current interactions in the form obtained in \cite{Gelfond:2010pm}. The
 same time, it elucidates  deep geometric structures underlying the perturbative
analysis of HS theory, relating homotopy integrals over different types of simplexes.
This field redefinition not only determines relative coefficients in
front of different current interactions but also suggests a proper criterium of
(non)locality in nonlinear $AdS_4$ HS theory, further
elaborated in \cite{loc} where it is shown in particular that the
results obtained in this paper are unambiguously selected by the proper locality criterion
of higher-order corrections.
 It should be stressed that the analysis of the 0-form sector of this paper
is fully informative implying locality in the 1-form sector up to possible gauge
artifacts. Details for the 1-form sector are presented in \cite{Prep}.

Note that the analysis of \cite{loc} refers mainly to the locality of the final result
that should not be confused with the issue of (non)locality of the field redefinition
from the originally nonlocal setup to the local one. The same time the results
of \cite{loc} showing that the field redefinition found in this paper is essentially
nonlocal are in agreement with the  analysis of
\cite{Taronna:2016xrm} where it was argued that
a field redefinition bringing the  nonlocal setup resulting from the standard
homotopy analysis of nonlinear equations to any local form is essentially nonlocal.
The interpretation of the results of this paper given in \cite{loc} provides
however a starting point for elaboration a perturbative scheme that operates
entirely with local or minimally nonlocal  results at higher orders with
no reference to nonlocal field redefinitions at all.

One of the  conclusions of this paper is that the current interactions in HS
theory are independent of the phase of the free parameter $\eta$ of the nonlinear HS
equations, depending solely on $\eta \bar \eta$.
 Remarkably, this  agrees with the  conjectures
on  holographic duality of the HS theory with different $\eta$
\cite{Aharony:2011jz,Giombi:2011kc} as well as with the structure of the boundary
correlator found in \cite{Maldacena:2012sf}.
Namely, following \cite{Vasiliev:2012vf} we demonstrate how different boundary
results originate from the $\eta$-dependent boundary conditions compatible with
the boundary CFT.

 The rest of the paper is organized as follows. In Section \ref{Free}
 we recall unfolded formulation of free massless fields and conserved
 currents in four dimensions. The structure of the unfolded equations
 describing current interactions of massless fields of all spins is
 recalled in Section \ref{deform}. The form of nonlinear HS equations is
 sketched in Section \ref{Nonlinear Higher-Spin Equations}.
 In Section \ref{Quadrat} quadratic
 corrections to massless equations resulting from the nonlinear HS equations
 are obtained  and their relation to usual local
 current interactions is established.  Holographic interpretation of the obtained
 results is  discussed in Section \ref{hol} with the emphasize on the dependence
 on the phase  parameter of the HS theory. Main results are summarized and discussed
 in Section \ref{disc}.

\section{Free massless equations}
\label{Free}
\subsection{Central on-shell theorem}
\label{CMT}
The infinite set of $4d$ massless fields of all spins $s=0,1/2,1,3/2,2\ldots $ is
conveniently described by a
{1-form} $  \omega (y,\bar{y}| x)= dx^n\omega_n (y,\bar{y}| x)$ and 0-form
$ C(y,\bar{y}| x)  $ \cite{Ann}
\be
\label{f}
 f(y,\bar{y}| x) =\f{1}{2i}\sum_{n,m=0}^{\infty} \frac{1}{n!m!}
{y}_{\alpha_1}\ldots {y}_{\alpha_n}{\bar{y}}_{{\pb}_1}\ldots
{\bar{y}}_{{\pb}_m } f{}^{\alpha_1\ldots\alpha_n}{}_,{}^{{\pb}_1
\ldots{\pb}_m}(x)\,,
\ee
where $x^n$ are $4d$ space-time coordinates and  $Y^A=(y^\ga, \bar y^\dga)$
are auxiliary commuting spinor variables ($A=1,\ldots 4$ is a Majorana spinor index
while $\ga = 1,2$ and $\dga =1,2$ are two-component spinor indices).
The key fact of the analysis of $4d$  massless fields
referred to as Central on-shell theorem is that unfolded system of field equations
for free massless
fields of all spins has the form \cite{Ann}
\bee
\label{CON1}
    && R_1(y,\overline{y}| x) = L(w,C):=
\f{i}{4} \Big ( \eta \overline{H}^{\dga\pb}\f{\p^2}{\p \overline{y}^{\dga} \p \overline{y}^{\dgb}}\
{\bar C}(0,\overline{y}| x) +\bar \eta H^{\ga\gb} \f{\p^2}{\p
{y}^{\ga} \p {y}^{\gb}}\
{C}(y,0| x)\Big )\,, \\\label{CON2}
\,&&  \D_{tw} C (y,\overline{y}| x) =0\,, \eee
where
\be
\label{RRR}
R_1 (y,\bar{y}| x) :=\D_{ad}\omega(y,{\bar{y}}|x) :=
D^L \omega (y,\bar{y}| x) +
 h^{\ga\pb}\Big (y_\ga \frac{\partial}{\partial \bar{y}^\pb}
+ \frac{\partial}{\partial {y}^\ga}\bar{y}_\pb\Big )
\omega (y,\bar{y} | x) \,,
\ee
\be
\label{tw}
 \D_{tw} C(y,{\bar{y}}|x) :=
D^L C (y,{\bar{y}}|x) -{i} h^{\ga\pb}
\Big (y_\ga \bar{y}_\pb -\frac{\partial^2}{\partial y^\ga
\partial \bar{y}^\pb}\Big ) C (y,{\bar{y}}|x)\,,
\ee
\be
\label{dlor}
D^L f (y,{\bar{y}}|x) :=
\dr f (y,{\bar{y}}|x) +
\Big (\go^{\ga\gb}y_\ga \frac{\partial}{\partial {y}^\gb} +
\overline{\go}^{\pa\pb}\bar{y}_\pa \frac{\partial}{\partial \bar{y}^\pb} \Big )
f (y,{\bar{y}}|x)\,.
\ee
Here $\dr =dx^n\f{\p}{\p x^n}$.
Background $AdS_4$ space
is described by a flat $sp(4)$
connection $w=(\omega_{\alpha \gb},\overline{\omega}_{\dga\dgb},h_{\ga\dgb})$
containing Lorentz connection
$\omega_{\alpha \gb} $, $\overline{\omega}_{\dga\dgb}$ and
vierbein  $h_{\ga\dgb}$ that obey the  equations
\be
\label{adsfl}
R_{\ga\gb}=0\,,\quad \overline{R}_{\pa\pb}=0\,,
\quad R_{\ga\pa}=0\,,
\ee
where, here and after discarding the wedge product symbols,
\be
\label{nR}
R_{\alpha \gb}:=\dr\omega_{\alpha \gb} +\omega_{\alpha}{}_\gamma
 \omega_{\gb}{}^{\gamma} -\, H_{\alpha \gb}\q
\overline{R}_{{\pa} {\pb}}
:=\dr\overline{\omega}_{{\pa}
{\pb}} +\overline{\omega}_{{\pa}}{}_{\dot{\gamma}}
 \overline{\omega}_{{\pb}}{}^{ \dot{\gga}} -\,
 \overline H_{{\pa\pb}}\,,
\ee
\begin{equation}
\label{nr}
R_{\alpha {\pb}} :=\dr  +\omega_\alpha{}_\gamma
h^{\gamma}{}_{\pb} +\overline{\omega}_{{\pb}}{}_{\dot{\delta}}
 h_{\alpha}{}^{\dot{\delta}}\,.
\end{equation}
Two-component indices are raised and lowered
by $\varepsilon_{\ga\gb}=-\varepsilon_{\gb\ga}$, $\varepsilon_{12}=1$: $A^\ga =\gvep^{\ga\gb} A_\gb$,
$A_\ga = A^\gb\gvep_{\gb\ga}$ and analogously for dotted indices.
$H^{\ga\gb}=H^{\gb\ga}$ and $\overline{H}^{\pa\pb} =
\overline{H}^{\pb\pa}$ are the frame 2-forms
\be
\label{H}
H^{\ga\gb} := h^{\ga}{}^\pa  h^\gb{}_\pa\,,\qquad
\overline {H}^{\pa\pb} := h^{\ga}{}^\pa h_{\ga}{}^{\pb}\,.
\ee
$\eta$ and $\bar \eta$ are complex conjugated free parameters introduced
for future convenience.
Their module will be identified with the coupling constant of the theory
\be
\label{nn}
\eta \bar \eta = f\,.
\ee
The phase of $\eta$ matters at the full nonlinear level as well as in
the holographic interpretation. The $AdS$ radius $\rho=\lambda^{-1}$ which is set equal to 1
in this paper can be reintroduced via rescaling $h_{\alpha{\pb}}\to \lambda h_{\alpha{\pb}}$.

1-form HS connection $  \omega (y,\bar{y}| x)$
contains HS gauge fields. For spins $s\geq 1$, equation (\ref{CON1})
expresses the Weyl {0-forms} $C(Y|x)$ via
gauge invariant combinations of derivatives of the HS gauge connections.
From this perspective the Weyl {0-forms} $C(Y|x)$ generalize the spin-two
Weyl tensor along with all its derivatives to any spin.

System (\ref{CON1}), (\ref{CON2}) decomposes into  subsystems
of different spins. In these terms, a massless spin $s$ is described by
the 1-forms $ \omega (y,\bar{y}| x)$ and 0-form $C (y,\bar{y}| x)$ obeying
\be
\omega (\mu y,\mu \bar{y}\mid x) = \mu^{2(s-1)} \omega (y,\bar{y}\mid x)\q
C (\mu y,\mu^{-1}\bar{y}\mid x) = \mu^{\pm 2 s}C (y,\bar{y}\mid x)\,,
\ee
where the signs  $+$ and $-$ correspond to self-dual and anti-self-dual parts
of the generalized Weyl tensors $C (y,\bar{y}| x)$ with helicities $h=\pm s$. More precisely,
$C(y,0| x)$ and $C(0,\bar y| x)$ describe the minimal gauge invariant combinations of derivatives of
the gauge fields of spins $s\geq 1 $ as well as the matter fields of spins $s=0$ or $1/2$.
For $s=1,2$, $C(y,0| x)$ and $C(0,\bar y| x)$ parameterize Maxwell and Weyl
tensors respectively. Those associated with higher powers of auxiliary variables
$y$ and $\bar y$ describe on-shell nontrivial combinations of derivatives
of the generalized Weyl tensors as is obvious from equations (\ref{CON2}), (\ref{tw})
relating second derivatives in $y,\bar y$ to the $x$ derivatives of  $C (Y|x)$ of lower
degrees in $Y$. Higher space-time derivatives in the nonlinear system result from the
 components of $C (Y|x)$ of higher degrees in $Y$.

Since twisted covariant derivative (\ref{tw})
contains both the term with two derivatives in $y$, $\bar y$ and that with the product of
$y$ and $\bar y$ the components of $C (y,\bar{y}| x)$
of the degree
\be
2N_x (C(Y|x)) := N_Y (C (Y|x)) - 2s\q
N_Y (C(Y|x)):= \big (y^\ga \f{\p}{\p y^\ga} + \bar y^\dga \f{\p}{\p\bar y^\dga}\big ) C(Y|x)
\ee
contain space-time derivatives of order $N_x (C(Y|x)$ and
lower. In other words, $AdS$ geometry induces filtration
with respect to space-time derivatives rather than  gradation as would be
 the case for massless fields in Minkowski space free of a dimensional parameter.
Because of the lack of indices, at any given order in $Y$, $C(Y|x)$ only mixes leading and first
subleading derivative of massless fields.

\subsection{Currents}

Conserved currents $J(Y_1,Y_2|x)$ are associated
 with  bilinears of the 0-forms $C(Y|x)$ \cite{Gelfond:2003vh,Gelfond:2006be}
\be
\label{JCC}
\PPP(Y_1,Y_2|x) := C(Y_1|x) \tilde C(Y_2|x)\q \tilde C (y,\bar y|x) =
C (-y,\bar y|x)\,.
\ee
The two types of fields $C (y,\bar y|x)$ and $\tilde C (y,\bar y|x)$ are analogues
of the fields $C_\pm(y,\bar y|x)$ of \cite{Gelfond:2010pm} characterized by
opposite signs  in the $h$-dependent term of  twisted covariant derivative
(\ref{tw}).

As a consequence of rank-one equation (\ref{tw})
$\PPP (Y_1,Y_2|x)$ obeys the rank-two equation
\be
\label{tw2}
\D^2_{tw} \PPP(Y_1,Y_2|x)=0 \q  \D_{tw}^2:=
D^L  -{i} h^{\ga\pb}
\Big (y_{1\ga} \bar{y}_{1\pb} - y_{2\ga} \bar{y}_{2\pb}-
\frac{\partial^2}{\partial y_1^\ga
\partial \bar{y_1}^\pb}+ \frac{\partial^2}{\partial y_2^\ga
\partial \bar{y_2}^\pb} \Big ) \,.
\ee
As shown in \cite{Gelfond:2006be,Gelfond:2010pm}, unfolded equation (\ref{tw2})
describes conserved conformal currents along with some
off-shell conformal currents in four dimensions.

The current $\PPP$ contains primary components $\PPP_0$, that cannot be expressed via
derivatives of others, and descendants $\PPP_+$ expressible via $x$-derivatives
of the  primaries. From (\ref{tw2}) it is obvious that
\be
J_0 \in Ker \,\gs_-\q
\gs_- = \frac{\partial^2}{\partial y_1^\ga
\partial \bar{y_1}^\pb}- \frac{\partial^2}{\partial y_2^\ga
\partial \bar{y_2}^\pb}\,.
\ee
Descendants $\PPP_+$  belong to a supplement
of $Ker \,\gs_-$. Being expressed via derivatives
of $\PPP_0$, descendant currents, most of which are also conserved,
correspond to  so-called improvements which do not generate
 nontrivial  charges. The space of $J_0$ was analyzed in
detail in \cite{Gelfond:2010pm,Gelfond:2015poa}.

At the free field level,  HS equations (\ref{CON1}), (\ref{CON2}) can be
supplemented with current conservation equation (\ref{tw2}). For a flat background connection
obeying (\ref{adsfl}) this unfolded system is  consistent
independently of whether $\PPP$ has  bilinear form (\ref{JCC}) or not.
(For instance, $\PPP$ can be realized by a sum over the color index $i$ of
$C^i(Y|x)$.) This leads to the remarkable fact emphasized in \cite{Gelfond:2010pm}
that the nonlinear equations describing current interactions in the lowest order
 amount to a linear system for $\go$, $C$ and $\PPP$.

\section{Current deformation}
\label{deform}
{Schematically,} {for a flat connection} $D=\dr+w$, $w=(\go,\bar \go, h)$ obeying
$(\D_{ad})^2=0$ (equivalently, $(\tilde \D_{tw})^2 =0$)
which is a concise form of (\ref{adsfl}), the current deformation
of  free equations (\ref{CON1}), (\ref{CON2}) has the form
  \be\label{defgo}
  \D_{tw}\go-L(w,C)+\G_{cur}(w ,\PPP)=0\,,
\ee
\be
\label{defc}  \D_{ad} C+\Hh_{cur}(w,\PPP)=0\,,
\ee
where $L(w,C)$ is defined in (\ref{CON1}) while
the 2-form $\G_{cur}(w ,\PPP)$ and 1-form $\Hh_{cur}(w,\PPP)$ are some functionals of the background fields $w$
and the current $\PPP$. Current interactions are associated with
$\G_{cur}(w ,\PPP)$ and $\Hh_{cur}(w,\PPP)$ linear in $\PPP$. Their form
 should respect the
consistency of (\ref{defgo}) and (\ref{defc}) with $\dr^2=0$ by virtue of current equation
(\ref{tw2}) that remains unchanged.

In this  setup  formal consistency of the system
is insensitive to whether  $\PPP$ is bilinear in the rank-one fields like
in (\ref{JCC}) or not. It is only important  that $J$ obeys
(\ref{tw2}). As a result,  system of equations (\ref{defgo}),
 (\ref{defc}) and (\ref{tw2}) is linear, admitting an interesting interpretation
 \cite{Gelfond:2010pm}
 as mixing massless fields in four dimensions associated with $\go$ and $C$
 and those in six dimensions associated with $\PPP$.  The same property implies a clear
 group-theoretical interpretation of equations (\ref{defgo}),
 (\ref{defc}) and (\ref{tw2}) as covariant constancy conditions in the
 appropriate deformation of the $o(3,2)\sim sp(4)$-modules  associated with
$\go$, $C$ and $\PPP$.

The functionals $\G_{cur}(w ,\PPP)$ and $\Hh_{cur}(w,\PPP)$ are  determined by the
compatibility conditions not uniquely. The freedom in
$\G_{cur}(w ,\PPP)$ and $\Hh_{cur}(w,\PPP)$
results from field redefinitions  linear in $\PPP$
\be
\label{fred}
\go \to \go' =\go + \Omega (w,\PPP)\q C\to C' = C+ \Phi (\PPP)\,.
\ee
 Deformations $\G_{cur}(w ,\PPP)$ and $\Hh_{cur}(w,\PPP)$ are nontrivial when they cannot be removed by
a field redefinition and trivial otherwise. Usual current interactions
are nontrivial. Trivial terms are associated with improvements. Schematically, we set
\be
\label{J}
J=J_0 +\Delta J\,,
\ee
where $\Delta J$ denotes an improvement that can be removed by a field redefinition
(\ref{fred}).

The problem  admits a cohomological interpretation with nontrivial
current interactions identified with the quotient of all
$\G_{cur}(w ,\PPP)$ and $\Hh_{cur}(w,\PPP)$ respecting compatibility over the trivial ones.
The resulting cohomology can be identified with certain Shevalley-Eilenberg cohomology
of $sp(4)$ with coefficients in $sp(4)$-modules associated with $\go$, $C$ and $\PPP$.
{Algebraically this is the semidirect sum  of a rank-one and rank-two systems.}
(For more detail on these issues we refer the reader to \cite{Gelfond:2015poa} and references therein.)
The cohomological
interpretation of the current interactions raises a problem of
 a representative choice equivalent to the choice of
variables modulo field redefinitions (\ref{fred}).

It should be stressed that the concept of (non)triviality
of the currents depends crucially on the choice of a proper class of
functions. Indeed, since the system in question is linear, it
admits a solution in the form
\be
\go(\PPP)  = \go_0 + G_\go(J)\q  C(\PPP)  = C_0 + G_C(J)\,,
\ee
where $G_\go$ and $G_C$ are appropriate Green functions while $\go_0$ and
$C_0$ solve the $\PPP$-independent part of the equations. The shift by a Green function
is a nonlocal operation. So,  currents can be nontrivial only with
respect to local field redefinitions. In the HS theory in $AdS$ space involving
 infinitely many spins and  derivatives,  the question  what is a proper
 substitute of the concept of a local field redefinition among various field redefinitions
(\ref{exp}), that preserves the concept of nontrivial current interactions, is not quite
trivial.

The current deformation of $4d$ HS field equations  was analyzed in
\cite{Gelfond:2010pm} where the local functionals $\G_{cur}(w ,\PPP)$ and $\Hh_{cur}(w,\PPP)$
 were found from the consistency conditions  in the basis where nontrivial
currents correspond to the primaries of the  current module $\PPP$. This corresponds to
$\G_{cur}(w ,\PPP)$ and $\Hh_{cur}(w,\PPP)$
in  (\ref{defgo}) and
(\ref{defc}) with the minimal number of derivatives which is finite for any
fixed spins $s_1$, $s_2$  of the constituent fields of the current $\PPP$ (\ref{JCC})
and the spin $s_\PPP$ of the current $\PPP$   equal to
 the spin of the fields $\go$ and $C$ in Eqs.~(\ref{defgo}), (\ref{defc}).

 The corresponding interactions classified by Metsaev \cite{Metsaev:2005ar} exist
 provided that for a conserved  currents with $s_\PPP\geq 1$
\be
\label{12j}
s_\PPP \geq s_1 +s_2\,
\ee
(Yukawa interaction with $s_\PPP, s_1, s_2=0$ or $1/2$  not involving conserved
currents does not obey this restriction).
Later on HS cubic interactions were considered by many authors in various setups
(see, e.g., \cite{Sagnotti:2010at}-\cite{Joung:2012hz} and also
\cite{Boulanger:2011dd, Sezgin:2011hq} for the nonstandard full action proposal).

In \cite{Gelfond:2010pm} it was shown that deformation (\ref{defgo}), (\ref{defc})
properly reproduces current interactions of HS fields. In the lowest
nontrivial order in interactions it is impossible to determine the coefficients
in front of individual currents solely from the consistency of the deformation.
As such they remain arbitrary functions of $s_\PPP$, $s_1$ and $s_2$.

For simplicity, in this paper we focus on the deformation $\Hh_{cur}(w,\PPP)$
in the 0-form sector, leaving consideration of $\G_{cur}(w,\PPP)$
to \cite{Prep}. Since $\Hh_{cur}(w,\PPP)$ and $\G_{cur}(w,\PPP)$ are tightly
related by the compatibility conditions,
analysis of the 0-form sector  answers all qualitative questions
on HS current interactions.
Upon the change of variables from $y^\pm$ used in \cite{Gelfond:2010pm} to
$y_{1,2}$ of this paper the final result of \cite{Gelfond:2010pm} in the 0-form
sector can be represented in the form
\bee
\label{F}
&&\Hh_{cur}(w,\PPP)= \f{1}{4}\int_0^1 d\tau\ls\,\sum_{h_1,h_2,h_\PPP}
\nn\\
&&
\ls\ls\ls\Big (a(h_1,h_2,h_\PPP)
\int\f{ d\bar s d\bar t}{(2\pi)^{2}} \exp i [ \bar s_\dgb  \bar t^\dgb ]
 h(y, \tau\bar s +(1-\tau) \bar t)
\PPP_{h_1,h_2,h_\PPP}(\tau y ; -(1-\tau) y  , \bar y +\bar s ; \bar y +\bar t)\nn\\
&&
\ls\ls\ls\ls+\bar a(h_1,h_2,h_\PPP)\int \f{d s d t}{(2\pi)^{2}}
 \exp i [  s_\gb   t^\gb ] h(\tau  s -(1-\tau) t,\bar y)
\PPP_{-h_1,-h_2,-h_\PPP}(y+s;y+t,\tau \bar y;- (1-\tau) \bar y )\Big )
\,,
\eee
where we use notation
\be
\label{h}
h(u,\bar u) = h^{\ga\dga} u_\ga \bar u_\dga\,
\ee
and $\PPP_{h_1,h_2,h_\PPP}$ is the projection of $\PPP$ to the helicities
${h_1,h_2,h_\PPP}$.
 Also we find it convenient to slightly re-order the arguments
putting first left and then right spinors
\be
\PPP(Y_1;Y_2 ) = \PPP (y_1;y_2,\bar y_1;\bar y_2)\,.
\ee

 The  coefficients $a(h_1,h_2,h_\PPP)$ and
$\bar a(h_1,h_2,h_\PPP)$ can be non-zero provided that  condition (\ref{12j}) is true.

Using that, for a polynomial
$P(y_1;y_2) $ of degrees $n_1$ and $n_2$ in $y_1$ and $y_2$, respectively,
\be
\int_0^1 d\tau P(\tau y_1\,; (1-\tau) y_2) = \f{n_1!n_2!}{(n_1+n_2 +1)!} P(y_1;y_2)\,,
\ee
Eq.(\ref{F}) just reproduces the coefficients of the deformation  of
 \cite{Gelfond:2010pm}.
In fact,
it is easier to check compatibility of the deformation  directly in the
form (\ref{F}) than in the component formalism of \cite{Gelfond:2010pm}.

Indeed, an elementary computation  shows that, skipping explicit reference to helicities,
\bee
\label{DF}
&& \D_{tw} \Hh_{cur}(w,\PPP)= -\f{1}{8}\int_0^1 d\tau \f{\p}{\p \tau}
\nn\\
&&\ls\ls\Big (a \int\f{ d\bar s d\bar t}{(2\pi)^{2}} \exp i [ \bar s_\dgb  \bar t^\dgb ]
\overline H^{\dga\dgb} (\tau \bar s_\dga + (1-\tau) \bar t_\dga)
(\tau \bar s_\dgb + (1-\tau) \bar t_\dgb)
\PPP(\tau y ; -(1-\tau) y  , \bar y +\bar s ; \bar y +\bar t)\nn\\
&&\ls\ls+\bar a \int\f{ d s d t}{(2\pi)^{2}} \exp i [  s_\gb  t^\gb ]
 H^{\ga\gb} (\tau  s_\ga + (1-\tau)  t_\ga)
(\tau  s_\gb + (1-\tau)  t_\gb)
\PPP(y+s; y+t,\tau \bar y ; -(1-\tau) \bar y)\Big )
\nn\\
&&
\ls\ls =-\f{1}{8}\Big (a
\int\f{ d\bar s d\bar t}{(2\pi)^{2}} \exp i [ \bar s_\dgb  \bar t^\dgb ]
\big ( \overline H^{\dga\dgb} \bar  s_\dga \bar s_\dgb  \PPP( y , 0  , \bar y +\bar s , \bar y +\bar t)
- \overline H^{\dga\dgb} \bar  t_\dga \bar t_\dgb \PPP( 0  ;-y , \bar y +\bar s ; \bar y +\bar t) \big)
\nn\\
&&
\ls\ls+\bar a\int \f{d s d t}{(2\pi)^{2}}
 \exp i [  s_\gb   t^\gb ]
\big (  H^{\ga\gb} s_\ga s_\gb \PPP( y+s ,y+t  , \bar y , 0)
- H^{\ga\gb} t_\ga t_\gb \PPP( y+s  ;y+t , 0; -\bar y ) \big )
\Big )\,.
\eee
 For appropriately
adjusted coefficients $a(h_1,h_2,h_\PPP) $ and $\bar a(h_1,h_2,h_\PPP) $
the remaining nonzero terms on
the {\it r.h.s} of (\ref{DF}) precisely cancel those produced by  differentiation
of $\go$ by virtue of
First on-shell theorem (\ref{CON1}) in the $\go$-dependent part of the
nonlinear deformation (\ref{HH}).
 However, as shown in \cite{Gelfond:2015poa},
such terms are nonzero only beyond the region (\ref{12j}). In other words,
$\Hh_{cur}$ (\ref{F}) is $ \D_{tw}$-closed for any coefficients provided that
(\ref{12j}) is true.

The deformation obtained in \cite{Gelfond:2010pm}
was  checked to give rise to proper current
interactions of different spins. This is also  easy to see  from  (\ref{F}).

Consider for instance the spin-one current. Let $h_\PPP=1$. In this case,
the primary sector
has degree 2 in $y$ and $0$ in $\bar y$. We observe that since $C(y,0|x)$
enters (\ref{CON1}) with the factor of
$\bar \eta$ the contribution of the current to the
Maxwell equations on the spin-one potential $\go(0,0|x)$ is
\bee
\label{F1}
&&{i} \bar \eta \Hh_{cur}(w,\PPP)\Big |_{h_\PPP = 1}= \f{i}{4}\bar\eta \int_0^1 d\tau\sum_{h_1,h_2}
a(h_1,h_2,1)
\int \f{ d\bar s d\bar t}{(2\pi)^{2}}\exp i [ \bar s_\dgb  \bar t^\dgb ] \nn\\
&&
\ls
 h(y, \tau\bar s +(1-\tau) \bar t)\Big (
\PPP_{h_1,h_2,1}(\tau y ; 0 , \bar s ; \bar t) +
\PPP_{h_1,h_2,1}(0; (1-\tau) y  , \bar s ; \bar t)\Big )
\,.
\eee

For $h_1=h_2=0$ this gives
\bee
{i}\bar \eta\Hh_{cur}(w,\PPP)\Big |_{h_1=h_2 =0,h_\PPP = 1}=&& \ls\f{i}{4}\bar \eta\, a(0,0,1)\int_0^1 d \tau \tau(1-\tau)
\int \f{ d\bar s d\bar t}{(2\pi)^{2}} \exp i [ \bar s_\dgb  \bar t^\dgb ]\nn\\&&
\ls\Big (h(y,  \bar t)
\PPP_{0,0,1}( y ; 0 , \bar s ;0 ) + h(y,  \bar s)
\PPP_{0,0,1}(0;  y  , 0 ; \bar t)\Big )
\eee
and, completing the integrations,
\be
\ls {i}\bar \eta\Hh_{cur}(w,\PPP)\Big |_{h_1=h_2 =0,h_\PPP = 1}= \f{\bar \eta}{24} a(0,0,1)
h(y,  \f{\p}{\p\bar t})\Big ( \PPP_{0,0,1}(y;0 ,  \bar t;0)-
\PPP_{0,0,1}(0;  y  , 0 ; \bar t)
\Big )_{ \bar t =0}\,.
\ee
Plugging in the bilinear expression (\ref{JCC}) for $\PPP$ and using rank-one
equation (\ref{CON2}) gives the usual spin-one current built from scalars,
$\PPP_n \sim \bar \eta C_1(0|x) \f{\p}{\p x^n} C_2 (0|x) - (1\leftrightarrow 2)$.
(For antisymmetrization over color indices one can start with matrix-valued
fields $C$.)

Analogously,  Eq.~(\ref{F1}) at $h_1=\half$ and $h_2 = -\half$
reproduces the canonical spin-one current built from spin-1/2 fields
\be
{i} \bar \eta\Hh_{cur}(w,\PPP)\Big |_{h_1=\half,h_2 =-\half,h_\PPP = 1}=
\f{\bar \eta}{12} a(\half,-\half,1)
h(y,  \f{\p}{\p\bar t}) \Big (\PPP_{-\half,\half,1}(0;  y  ,  \bar t; 0)
-\PPP_{\half,-\half,1}(y;0 , 0; \bar t)
\Big )_{ \bar t =0}\,.
\ee

Also it is  easy to see that formula (\ref{F}) properly reproduces
Yukawa interactions in the sector of spins 0 and 1/2 while the selfinteraction
of scalars is absent in agreement with the conclusion of \cite{Gelfond:2015poa}
that HS interactions in the 0-form sector are conformally invariant,
as well as with   the
observation that cubic coupling of scalars in the nonlinear HS theory
is zero \cite{Sezgin:2003pt}.

Let us stress that no nontrivial deformations beyond usual current interactions
can be expected in the lowest order because conserved currents fill in all
$\gs_-$ cohomologies in free unfolded HS equations (\ref{CON1}), (\ref{CON2}),
\ie there is no room for further nontrivial interactions.
(For more detail on the $\gs_-$ cohomology technics  see e.g. \cite{V_obz3}
and references therein.)

As mentioned above a
trivial part of the current $\PPP$ can be removed by a (generalized) local field redefinition while the
nontrivial part survives.
However, there is a subtlety. Since HS theories are formulated in $AdS$ space
of  non-zero radius $\rho=\lambda^{-1}$,  nonlinear HS equations
can give rise to infinite tails of higher-derivative
terms. It is neither guaranteed nor needed that
 general conserved currents   admit an expansion
in power series of individual improvements
$J=\sum_{n=0}^\infty a_n \rho^n J_n$ where $J_0$ is a spin-$s$ primary current
 built from some constituent fields $C$, while $J_n$ are descendants
containing up to $n$ derivatives of $J_0$.
  What is really necessary is that the current $J$, that appears
on the {\it r.h.s} of the field equations, should allow a representation (\ref{J})
with $\Delta J$ removable by a generalized local field redefinition.

A nonlinear deformation following
from nonlinear HS equations reproduces   current interactions
(\ref{F}) with certain coefficients $a(h_1,h_1,h_\PPP)$.
Before going into detail of their derivation in Section \ref{Quadrat},
we briefly sketch the form of nonlinear  HS equations.

\section{Nonlinear higher-spin equations in $AdS_4$}
\label{Nonlinear Higher-Spin Equations}

The key element of the construction of \cite{more} is
the dependence of the  HS 1-forms and 0-forms
 on an additional Majorana spinor variable $Z^A$ and
Klein operators $K=(k,\bar{k})$
\be
\go(Y;K|x)\longrightarrow W(Z;Y;K|x)\,,\qquad
C(Y;K|x)\longrightarrow B(Z;Y;K|x)\,.
\ee

An additional spinor field $S_A (Z;Y;K|x)$, that carries only pure gauge
degrees of freedom, plays a role of connection in
 $Z^A$ directions. It is convenient to
introduce anticommuting $Z-$differentials $\theta^A$, $\theta^A \theta^B=-\theta^B
\theta^A$, to interpret $S_A (Z;Y;K|x)$ as a 1-form in $Z$ direction,
\be
S=\theta^A S_A (Z;Y;K|x) \,.
\ee

 HS equations determining  dependence on the
 variables $Z_A$ in terms of ``initial data"
\be
\label{inda}
\go(Y;K|x)=W(0;Y;K|x)\,,\qquad C(Y;K|x)= B(0;Y;K|x)\,,
\ee
 are formulated in terms of the
associative star product $*$ acting on functions of two
spinor variables
\be
\label{star2}
(f*g)(Z;Y)=
\int \f{d^{4} U\,d^{4}V}{(2\pi)^{4}}  \exp{[iU^A V^B C_{AB}]}\, f(Z+U;Y+U)
g(Z-V;Y+V) \,,
\ee
where
$C_{AB}=(\epsilon_{\ga\gb}, \bar \epsilon_{\dga\dgb})$
is the $4d$ charge conjugation matrix and
$ U^A $, $ V^B $ are real integration variables.
 1 is a unit element of the star-product
algebra, \ie $f*1 = 1*f =f\,.$ Star product
(\ref{star2}) provides a particular
realization of the Weyl algebra
\be
[Y_A,Y_B]_*=-[Z_A,Z_B ]_*=2iC_{AB}\,,\qquad
[Y_A,Z_B]_*=0\q [a,b]_*:=a*b-b*a\,.
\ee

The  Klein operators $K=(k,\bar k)$ satisfy
\bee
\label{kk}
&&k* w^\ga = -w^\ga *k\,,\quad
k * \bar w^\pa = \bar w^\pa *k\,,\quad
\bar k *w^\ga = w^\ga *\bar k\,,\quad
\bar k *\bar w^\pa = -\bar w^\pa *\bar k\,, \\  && k*k=\bar k*\bar k = 1\,,\quad
k*\bar k = \bar k* k\,
\eee
 with
 $w^\ga= (y^\ga, z^\ga, \theta^\ga )$, $\bar w^\pa =
(\bar y^\pa, \bar z^\pa, \bar \theta^\pa )$. These relations extend the action of the star product to the
Klein operators.

The nonlinear HS equations are \cite{more}
\be
\label{dW}
\dr W+W*W=0\,,\qquad
\ee
\be
\label{dB}
\dr B+W*B-B*W=0\,,\qquad
\ee
\be
\label{dS}
\dr S+W*S+S*W=0\,,
\ee
\be
\label{SB}
S*B=B*S\,,
\ee
\be
\label{SS}
S*S= i (\theta^A \theta_A + \theta^\ga \theta_\ga  F_*(B)* k*\kappa +
\bar \theta^\dga \bar \theta_\dga \bar F_*(B)* \bar k* \bu)
\,,
\ee
where
$F_*(B) $ is some star-product function of the field $B$.
The simplest choice  of linear functions
\be
\label{etaB}
F_*(B)=\eta B \q \bar F_* (B) = \bar\eta B\,,
\ee
where $\eta$ is a complex parameter
\be
\label{theta}
\eta = |\eta |\exp{i\varphi}\q \varphi \in [0,\pi)\,,
\ee
leads to a class of pairwise nonequivalent nonlinear HS
theories. The  cases
of $\varphi=0$ and $\varphi =\f{\pi}{2}$  correspond
to so called $A$ and $B$ HS models distinguished
by the property that they  respect parity \cite{Sezgin:2003pt}.

The left and right inner Klein operators
\be
\label{kk4}
\kappa :=\exp i z_\ga y^\ga\,,\qquad
\bu :=\exp i \bar{z}_\dga \bar{y}^\dga\,,
\ee
 which enter Eq.~(\ref{SS}), change a sign of
 undotted and dotted spinors, respectively,
\be
\label{uf}
\!(\kappa *f)(z,\!\bar{z};y,\!\bar{y})\!=\!\exp{i z_\ga y^\ga }\,\!
f(y,\!\bar{z};z,\!\bar{y}) ,\quad\! (\bu
*f)(z,\!\bar{z};y,\!\bar{y})\!=\!\exp{i \bar{z}_\dga \bar{y}^\dga
}\,\! f(z,\!\bar{y};y,\!\bar{z}) ,
\ee
\be
\label{[uf]}
\kappa *f(z,\bar{z};y,\bar{y})=f(-z,\bar{z};-y,\bar{y})*\kappa\,,\quad
\bu *f(z,\bar{z};y,\bar{y})=f(z,-\bar{z};y,-\bar{y})*\bu\,,
\ee
\be
\kappa *\kappa =\bu *\bu =1\q \kappa *\bu = \bu*\kappa\,.
\ee

Perturbative analysis of  Eqs.~(\ref{dW})-(\ref{SS}) assumes their
linearization around some vacuum solution.
The simplest choice is
\be
\label{vacs}
W_0(Z;Y;K|x)= w(Y|x)\q S_0(Z;Y;K|x) = \theta^A Z_A\q B_0(Z;Y;K|x)=0\,,
\ee
where $w(Y|x)$ is some solution to the flatness condition
$\dr w + w*w=0$.
A flat connection $w(Y|x)$ bilinear in $Y^A$ describes $AdS_4$
\be
\label{ads}
w(Y|x) = -\f{i}{4} w^{AB} Y_A Y_B = -\f{i}{4} (\go^{AB} + h^{AB}) Y_A Y_B\,,
\ee
\be
\go^{AB} Y_A Y_B  := \go^{\ga\gb} y_\ga y_\gb + \bar \go^{\dga\dgb}
\bar y_\dga \bar y_\dgb \q h^{AB} Y_A Y_B  := 2h^{\ga\dgb} y_\ga \bar y_\dgb\,.
\ee

Propagating massless fields are described by $K$-even
$W(Z;Y;K|x)$ and $K$-odd $B(Z;Y;K|x)$
\be
W(Z;Y;-K|x)=W(Z;Y;K|x)\q B(Z;Y;-K|x)=-B(Z;Y;K|x)\,.
\ee
The fields of opposite parity in the Klein operators
\be
W(Z;Y;-K|x)=-\W(Z;Y;K|x)\q B(Z;Y;-K|x)=B(Z;Y;K|x)\,
\ee
are topological in the sense that every irreducible field describes at
most a finite number of degrees of freedom. (For more detail see
\cite{more,Vasiliev:1999ba,Didenko:2014dwa}). They can be
treated as describing infinite sets of  coupling constants in HS
theory. In this paper all these fields are set to zero.

It is easy to see  that in the first nontrivial order
\be
\label{B1C}
B_1(Z;Y;K|x) = C(Y;K|x)
\ee
(for more detail see \cite{more,Vasiliev:1999ba,Didenko:2014dwa}).
Thus, nonlinear HS equations (\ref{dW})-(\ref{SS}) give rise to
 the doubled set of massless fields
\be
\label{cij}
C(Y;K|x)= C^{1,0}(Y|x) k + C^{0,1}(Y|x) \bar k\,.
\ee

In the sector of massless fields, the linearization of Eqs.~(\ref{dW})-(\ref{SS})
 just reproduces free  equations (\ref{CON1}), (\ref{CON2})
\cite{more,Vasiliev:1999ba,Didenko:2014dwa} (see \cite{Didenko:2015cwv} for a simple derivation)
\bee
\label{CON1k}
    && \ls\ls \ls D_0 \go(y,\overline{y};K|x) =
\frac{i}{4} \left ( \eta \overline{H}^{\dga\pb}\f{\p^2}{\p \overline{y}^{\dga} \p \overline{y}^{\dgb}}\
{C}(0,\overline{y};K| x)k +\bar \eta H^{\ga\gb} \f{\p^2}{\p
{y}^{\ga} \p {y}^{\gb}}\
{C}(y,0;K| x)\bar k \right ), \\\label{CON2k}
\,&& \ls\ls {D}_0 C (y,\overline{y};K| x) =0\,,
\eee
where
\be
D_0 \go := \dr \go +w*\go +\go*w\q D_0 C:=\dr C +w*C -C*w\,.
\ee
The twisted adjoint covariant derivative acting on $C^{i,1-i}(Y|x)$ in
(\ref{CON2}) results from the application of the adjoint covariant derivative
to $C(Y;K|x)$ in (\ref{CON2k}) by virtue of (\ref{kk}).

For the analysis of HS field equations it is useful to unify the 1-forms $W$ and $S$
into a single field
\be
\label{sfield}
\W (Z;Y;K;\theta|x) =\dr_x + W (Z;Y;K|x) + S(Z;Y;K;\theta|x)\,,
\ee
where the space-time differentials $dx^n$ are implicit. In these terms, system
(\ref{dW})-(\ref{SS}) acquires the concise form
\be
\W*\W= i (\theta^A \theta_A + \theta^\ga \theta_\ga  F_*(B)* k*\kappa +
\bar \theta^\dga \bar
\theta_\dga \bar F_*(B)* \bar k *\bu)
\,,
\ee
\be
[\W\,, B]_* =0\,.
\ee

\section{Quadratic corrections in the 0-form sector}
\label{Quadrat}

\subsection{Quadratic deformation from nonlinear higher-spin equations}

First nontrivial correction to equations on the 0-forms $C$
has the form
\be
\label{HH}
 D_0 C+ [\go\,, C]_*+ \Hh (w,\PPP)=0\,,
\ee
where $\go$ stands for the first-order (\ie  not containing the
vacuum part $w$) part of  the $Z$-independent part of HS connection
(\ref{inda})  and $\PPP  $ denotes bilinears of the 0-forms $C$
\be \label{cur}
\PPP (y_1;y_2,\bar y_1;\bar y_2;K
|x) := C(y_1,\bar y_1;K|x)
 C(y_2,\bar y_2;K|x)\,.
\ee
The presence of either $k$ or $\bar k$ in the first factor of $C$ leads to
the change of a relative sign between the sector of $Y_1$ and $Y_2$
as in covariant derivative (\ref{tw2}) upon the Klein operator in the first factor of
$C(y_1,\bar y_1;K|x)$ is moved to the right through the second factor.

 In these terms,
\be
\Hh (w,\PPP) = \Hh_\eta (w,\PPP)+\Hh_{\bar{\eta}} (w,\PPP)\,,
\ee
where
\bee
\label{CC}
\Hh_\eta (w,\PPP) =&&\ls-\f{i}{2}  \eta\int\f{ dS dT}{(2\pi)^{4}}
\exp i S_A T^A \int^1_0 d\tau \nn\\
&& [ h(s, \tau \bar y - (1-\tau) \bar t)
\PPP( \tau s;-(1-\tau)y +t, \bar y +\bar s;\bar y+\bar t;K)\nn\\
&&- h (t, \tau \bar y  - (1-\tau)\bar s) \PPP
((1-\tau)  y +  s; \tau  t, \bar y+ \bar s;\bar y+\bar t;K)]*k \,,
\eee
\bee
\label{barCC}
\Hh_{\bar\eta} (w,\PPP) =&&\ls-\f{i}{2}  \bar \eta \int \f{dS dT}{(2\pi)^{4}}
\exp i S_A T^A \int^1_0 d\tau
\nn\\
&&\ls\ls[  h (\tau  y  - (1-\tau) t,\bar s) \PPP
(y+s;y+t, \tau \bar s; -(1-\tau)\bar y +\bar t;K)\nn\\
&&\ls\ls- h ( \tau  y  - (1-\tau) s,\bar t) \PPP
(y+s;y+t,(1-\tau) \bar y + \bar s; \tau \bar t);K]\bar *k\,.
\eee

 Formulae (\ref{CC}), (\ref{barCC}) are simple consequences of the homotopy formulae
obtained in \cite{Didenko:2015cwv}
\be
D_0 C + \HH_{tw} ([ \W_1 (Z;Y;K;\theta|x)\,,C(Y;K|x)]_*)=0\,,
\ee
where
\begin{eqnarray}
\qquad\HH_{tw} f(Z;Y;K|x) &&\ls :=
\exp\left\{ -\dfrac{i}{8}\omega^{AB}h_{A}{}^{C}\dfrac{\partial^{2}}{\partial
\theta^{B}\partial \theta^{C}}+\dfrac{i}{2} h^{AB}Y_{A}\dfrac{\partial}{\partial \theta^{B}}\right\} \nonumber \\
 &  & \ls\ls\ls\ls\exp\left\{-\dfrac{1}{2} \omega^{AB}\dfrac{\partial^{2}}{\partial Y^{A}\partial
 \theta^{B}}+\dfrac{1}{4}h^{AB}\dfrac{\partial^{2}}{\partial Z^{A}\partial \theta^{B}}\right\}
f(Z;Y;K;\theta|x) \Big |_{Z=\theta=0}
\label{eq:Htw}
\end{eqnarray}
and $\W_1 (Z;Y;K;\theta|x)$ is the first-order  correction to $\W$
which, using  (\ref{B1C}), has the form
\be
\W_1 (Z;Y;K;\theta|x) = \go(Y;K|x)  +i
\Delta_{ad}^{*}( \theta^\ga \theta_\ga  \eta\, C *\kappa k+
\bar \theta ^\dga \bar
\theta_\dga \bar  \eta\, C *\bu  \bar k)
\,,
\ee
where, as shown in \cite{Didenko:2015cwv},
\begin{equation}
\label{inte}
\Delta_{ad}^{*}f (Z;Y;\theta)=\dfrac{i}{2}Z^{A}\dfrac{\partial}{\partial \theta^{A}}
\int_{0}^{1}\frac{d\tau}{\tau} \exp \Big ( - \frac{1-\tau}{2\tau}w^{B C}
\f{\p^2}{\p Y^B \ \p\theta^C} \Big)
f (\tau Z;Y;\tau \theta)\,.
\end{equation}

In a slightly different form $\Hh (w,\PPP) $ was also
presented in \cite{Boulanger:2015ova}.

Being derived from the consistent nonlinear equations, deformation terms
(\ref{CC}) and (\ref{barCC}) obey the compatibility conditions.  Analogously to the analysis of the
deformation in Section \ref{deform} one can check that
\be
 D_0 \Hh_\eta (w,\PPP)=
\int_0^1 d \tau \f{\p}{\p \tau} G(\tau) = G(1)-G(0)\,,
\ee
where $G(1)=0$ because of the cancellation between the two terms in (\ref{CC}) while
$G(0)$ has the structure analogous to the {\it r.h.s.} of (\ref{DF}) canceling the
terms resulting from the differentiation of $\go$ in $[\go\,, C]$
by virtue of (\ref{CON1}).

The integration over $S$ and $T$ in (\ref{CC}), (\ref{barCC})  brings infinite tails of contracted indices.
As explained in Section \ref{CMT} this induces
an infinite expansion in higher space-time derivatives of the constituent
fields. Hence, formula (\ref{HH}) with $\Hh$ (\ref{CC}), (\ref{barCC}) differs from the conventional
current interactions  (\ref{F}) which, being free of the integration over $s_\ga$
and $t_\ga$, contains a finite number of derivatives for any given spins $s_1$, $s_2$ of
the constituent fields and $s_J$ of the current.

To reproduce standard current interactions from nonlinear HS equations
we have to find a field redefinition
\be
\label{red}
C \to C'(Y;K|x) = C(Y;K|x)+ \Phi (Y;K|x)
\ee
with $\Phi$  linear in $\PPP$, bringing equation (\ref{HH}) to the form
(\ref{defc}), (\ref{F})  with some coefficients $a(s_1,s_2,s_\PPP)$.
Before going into details of its derivation  note that, being
somewhat analogous, formulae (\ref{F}) and (\ref{CC}), (\ref{barCC}) are essentially different.
Namely, comparing the terms that contain usual star product
with respect to the barred variables we observe that the arguments
of the vierbein $h$ depend on $\bar y$ in (\ref{CC}) but on $y$ in (\ref{F}).
Naively, this makes it difficult to relate the two formulae by a field redefinition.
Nevertheless, as shown below this can be achieved in two steps. Firstly we find
in Section \ref{exec}
a field redefinition which, bringing the current terms to a local form, completely
removes the dependence on $y$ and $\bar y$ from the arguments of $h$. Secondly,
in Section \ref{reinc}, the deformation is brought by a local field redefinition
to the canonical form (\ref{F}) where $h$ depends on $y$ and $\bar y$ of opposite
chiralities compared to (\ref{CC}), (\ref{barCC}).

This chirality flip has a dramatic effect implying that the effective coupling constant
in front of currents depends on $\eta\bar\eta$, hence
 resolving some paradoxical claims in the literature  on the
meaningless dependence on the parameter $\eta$  (see
\cite{Boulanger:2015ova,Skvortsov:2015lja} and references therein).
More precisely the authors of \cite{Boulanger:2015ova,Skvortsov:2015lja}
were assuming that the current deformation results from the sectors $\eta^2$
and $\bar\eta^2$. If true, this would imply that the current contribution to the r.h.s.
of the Fronsdal equations is not sign definite for an arbitrary phase of $\eta$, changing
its sign under the phase rotation $\eta\to i\eta$, $\bar \eta\to -i\bar\eta$.
In particular, this would imply that the coefficient in front of the stress
tensor, \ie the gravitational constant, is not positive definite that would indicate
a serious inconsistency of either  the applied approach or  the field equations in
question. Our results show that the  approach  proposed in this paper is free of this problem.

\subsection{Ansatz}
\label{ans}

An appropriate Ansatz for the first field redefinition is
\be
\label{phi1}
\Phi_{1\eta} (Y;K|x) = \int \f{dS dT}{ (2\pi)^4}  \exp i S_A T^A \int \prod_{i=1}^3
d\tau_i \phi_{1\eta}(\tau_i)
\f{\p}{\p \tau_3}
\PPP( \tau_3 s+\tau_1y;t-\tau_2 y , \bar y +\bar s;\bar y+\bar t;K)*k\,,
\ee
where $\phi_{1\eta}(\tau_i)$ is some function of  three integration variables $\tau_i$.
The $\tau$-integration is over $\mathbb{R}^3$ while
$\phi_{1\eta}(\tau)$ is demanded to have a compact support allowing to
freely integrate by parts.

An important feature of the Ansatz (\ref{phi1}) is that the right sector in
(\ref{phi1}) is of the  form  of the star product with respect to dotted spinors, involving no homotopy parameters.  The rationale behind this
Ansatz is the usual separation variables with respect to $z_\ga$ and $\bar z_\dga$ sectors
 expressed by the two facts. The first is that the action of the inner Klein operators $\kappa$ and
 $\bar {\kappa}$ in the $F$ and $\bar F$-dependent terms on
 the \rhs of (\ref{SS}) leaves, respectively,  the right and left sectors untouched.
 The second is
 that the vacuum part $S_0$ (\ref{vacs}) is a sum of mutually commuting left and right components
\be
S_0=s_0 +\bar s_0\q s_0 := \theta^\ga z_\ga\q \bar s_0 := \bar\theta^\dga \bar z_\dga\,.
\ee

Reconstruction of nonlinear corrections to HS equations is based on the
perturbative resolution for
the dependence on $Z$ variables from the equations (\ref{dS}), (\ref{SB}) and (\ref{SS})
using the fact that the star-commutators with $s_0$ and $\bar s_0$ are proportional
to de Rham derivatives in the left and right sectors of $z_\ga$ and $\bar z_\dga$,
respectively. Following  the standard separation of
variables approach, it is  natural to look for a proper resolution operator $\dr^*_Z$ for $S_0$
 in the factorized form
\be
\dr^*_Z=\dr^*_z+\dr^*_{\bar z}\,
\ee
with independent left and right resolutions $\dr^*_z$ and  $\dr^*_{\bar z}$.
This  leads  to the factorized Ansatz (\ref{phi1}) for any $\dr^*_z$ and  $\dr^*_{\bar z}$.
Under this factorization condition  Eq.~(\ref{phi1}) can be checked to describe
the only field  redefinition that gives rise to local quadratic corrections to the
HS equations.  The details of this analysis are not presented in this paper since it
is elementary to prove the uniqueness  using the Green function obtained
in \cite{loc} where it is also  shown that the solution presented in this paper
plays a distinguished role from the locality perspective,
providing a proper local frame in the HS theory.

\subsection{Elimination of $\bar y$}
\label{exec}

An elementary  computation using the identity
\be
\label{ident}
\int
\f{ds dt}{(2\pi)^{2}} \exp i [ s_\ga  t^\ga ]
\Big(y_\ga \f{\p}{\p \tau_3} - i \big ( \p_{2\ga}\f{\p}{\p \tau_1} +
\p_{1\ga}\f{\p}{\p \tau_2}\big)\Big)
\PPP( \tau_3 s+\tau_1y;-\tau_2 y +t, \bar y +\bar s;\bar y+\bar t;K)=0\,,
\ee
 expressing the fact that antisymmetrization over any three two-component indices
is zero, yields
\bee
\label{rel1}
 D_0 \Phi_{1\eta}(Y;K|x) = &&\ls i h^{\ga\dga}\int
\f{dS dT}{(2\pi)^{4}} \exp i [ S_A  T^A ]
\int d^3\tau_i \phi_{1\eta}(\tau)\nn\\ &&\ls\ls\ls\ls\Big [
 (\tau_2 \bar \p_{1\dgb} -\tau_1 \bar \p_{2\dgb})(\p_{2\ga} \f{\p}{\p \tau_1}+
\p_{1\ga} \f{\p}{\p \tau_2}) -i (1-\tau_1-\tau_2)
(\p_{2\ga} \f{\p}{\p \tau_1} +
\p_{1\ga} \f{\p}{\p \tau_2})\bar y_\dgb
\nn\\&&  \ls+\f{\p}{\p \tau_3}\Big ( i\tau_3
(\p_{2\ga} + \p_{1\ga})
\bar y_\dgb -(1-\tau_1-\tau_3)\p_{1\ga}\bar \p_{1\dgb} + (1-\tau_2-\tau_3)
\p_{2\ga}\bar \p_{2\dgb} \nn\\&&\ls\ls+ \tau_1 \p_{1\ga}\bar \p_{2\dgb}-\tau_2
 \p_{2\ga}\bar \p_{1\dgb}
\Big )\Big ]
\PPP( \tau_3 s+\tau_1y;-\tau_2 y +t, \bar y +\bar s;\bar y+\bar t;K)*k\,.
\eee
Here $\p_{1\ga}$($\bar \p_{1\dga}$) and $\p_{2\ga}$($\bar \p_{2\dga}$)
denote derivatives over the first and second  undotted(dotted)
spinorial arguments of $J$ defined to anticommute with the respective Klein operators
$k(\bar k)$ so that
\be
\p_2 A(w_1)k B(w_2) = - A(w_1)k\p_2 B(w_2)\,.
\ee
This sign change matters in  relations (\ref{ident}), (\ref{rel1}) with the current $\PPP$
(\ref{cur}) built from two 0-forms $C$ containing $k$ or $\bar k$.

Setting
\be
\label{3t}
\phi_{1\eta}(\tau) = \phi_{1\eta}'(\tau) \delta\left ( 1-\sum_{i=1}^3 \tau_i \right )
\ee
and integrating by parts we obtain
\bee
\label{bphi}
&&\ls D_0 \Phi_{1\eta} (Y;K|x) =  i\int \f{dS dT}{(2\pi)^{4}}  \exp i [ S_A  T^A ]
\int d^3\tau \times \phi_{1\eta}'(\tau) \delta( 1-\sum_{i=1}^3 \tau_i )\nn\\
&&
\Big \{i h( \p_2, \bar y) \Big ( \f{\overleftarrow{\p}}{\p \tau_1}
- \f{\overleftarrow{\p}}{\p \tau_3}\Big )\tau_3
+i h( \p_1, \bar y)\Big (\f{\overleftarrow{\p}}{\p \tau_2}- \f{\overleftarrow{\p}}{\p \tau_3}\Big )\tau_3
-h( \p_1, \bar \p_1) \Big ( \f{\overleftarrow{\p}}{\p \tau_2}
 - \f{\overleftarrow{\p}}{\p \tau_3}\Big )\tau_2\nn\\
&&
+ h( \p_2, \bar \p_2)\Big (\f{\overleftarrow{\p}}{\p \tau_1}
- \f{\overleftarrow{\p}}{\p \tau_3}\Big )\tau_1
-h( \p_1, \bar \p_2) \Big ( \f{\overleftarrow{\p}}{\p \tau_3}
- \f{\overleftarrow{\p}}{\p \tau_2}\Big )\tau_1
+ h( \p_2, \bar \p_1)\Big (\f{\overleftarrow{\p}}{\p \tau_3}
- \f{\overleftarrow{\p}}{\p \tau_1}\Big )\tau_2\Big \}\nn\\
&&
\PPP( \tau_3 s+\tau_1y;-\tau_2 y +t, \bar y +\bar s;\bar y+\bar t;K)*k\,.
\eee
Finally, for
\be
\phi'_{1\eta} =\half  \eta\theta(\tau_1)\theta(\tau_2)\theta(\tau_3)\,,
\ee
this gives
\bee
 D_0 \Phi_{1\eta} (Y;K|x) = &&\ls-\frac{i}{2}\eta\int
\f{dS dT}{(2\pi)^{4}}  \exp i [ S_A  T^A ]
\int_0^1 d \tau\nn\\
&& \Big [h(s,\tau \bar y - (1-\tau)\bar t)\PPP( \tau s;-(1-\tau) y +t, \bar y +\bar s;\bar y+\bar t)\nn\\
&& -h(t,\tau \bar y - (1-\tau)\bar s)\PPP(  s+(1-\tau) y; \tau t, \bar y +\bar s;\bar y+
\bar t)\nn\\
&&\ls\ls -ih(\p_1+\p_2,(1-\tau)\bar t +\tau \bar s)
\PPP( \tau y;-(1-\tau) y , \bar y +\bar s;\bar y+\bar t;K)\Big ]*k\,.
\eee
Comparing this with (\ref{CC}) we find that
\be
\Hh_\eta (w,\PPP) =   D_0 \Phi_{1\eta}(J) + \Hh_\eta ' (w,\PPP)\,,
\ee
where, completing integration over $s$ and $t$,
\be
\label{hp}
\Hh _\eta' (w,\PPP)= \frac{\eta}{2} \int \f{d\bar s d\bar t}{(2\pi)^{2}}
\exp i [ \bar s_\dga \bar t^\dga ]
\int_0^1 d\tau
h(\p_1+\p_2,(1-\tau)\bar t +\tau \bar s)
\PPP( \tau y;-(1-\tau) y , \bar y +\bar s;\bar y+\bar t;K)*k\,.
\ee

We observe that, being free of the integration over $s$ and $t$,
$\Hh_\eta ' (w,\PPP)$ is a local functional of $\PPP$.
However,  containing a finite number of derivatives
of physical (\ie primary) fields for any three spins $s_1$, $s_2$ and $s_\PPP$,
 $\Hh_\eta ' (w,\PPP)$ differs from $\Hh_{cur}$ (\ref{F}).
Depending on $\p_1 + \p_2$ instead of $y$, $\Hh_\eta ' (w,\PPP)$ contains
one extra space-time derivative compared to (\ref{F}).
In Minkowski space this would imply that $\Hh_\eta ' (w,\PPP)$ was an improvement.
However this is not the case in $AdS_4$ where derivatives of different
orders talk to each other via nonzero commutators of covariant derivatives.

\subsection{Reincarnation of $y$}
\label{reinc}
Let
\be
\label{phi2}
\Phi_{2\eta}= -\frac{1}{2} \eta \int \f{d\bar s d\bar t}{(2\pi)^{2}}  \exp i [ \bar s_\dgb  \bar t^\dgb ]
\int_0^1 d\tau
\PPP(\tau y; -(1-\tau) y , \bar y +\bar s ; \bar y +\bar t; K)*k
\,.
\ee
An elementary computation yields
\be
D_0 \Phi_{2\eta}= -\frac{i}{2}\eta
  \int \f{d\bar s d\bar t}{(2\pi)^{2}}  \exp i [ \bar s_\dgb  \bar t^\dgb ]
\int_0^1 d\tau h(y -i(\p_2+\p_1), \tau\bar s +(1-\tau) \bar t)
\PPP(\tau y ; -(1-\tau) y  , \bar y +\bar s ; \bar y +\bar t;K)*k \,.
\ee
This gives
\be
\Hh_{\eta} (w,\PPP) =  \Hh_\eta{}_{\,impr}  (w,\PPP)
+ \Hh_\eta {}_{\,cur} (w,\PPP)\,,
\ee
where
\be
\Hh_\eta {}_{\,impr}  (w,\PPP):=  D_0 (\Phi_{1\eta} +\Phi_{2\eta})
\ee
and
\be
\label{hh}
\Hh_{\eta \,cur} (w,\PPP)= -\frac{i}{2}\eta
\int \f{d\bar s d\bar t}{(2\pi)^{2}}  \exp i [ \bar s_\dgb  \bar t^\dgb ]
\int_0^1 d\tau h(y, \tau\bar s +(1-\tau) \bar t)
\PPP(\tau y ; -(1-\tau) y  , \bar y +\bar s ; \bar y +\bar t;K) *k\,.
\ee
Analogously, in the $\bar \eta$ sector
\be
\label{barhh}
\Hh_{\bar \eta\,cur}  (w,\PPP)=  -\frac{i}{2}{\bar \eta}\int
\f{d s d t}{(2\pi)^{2}}  \exp i [  s_\gb   t^\gb ]
\int_0^1 d\tau h( \tau s +(1-\tau)  t,\bar y)
\PPP(   y + s ;  y + t,\tau \bar y ; -(1-\tau) \bar y;K ) *\bar k\,.
\ee
This expressions just reproduce (\ref{F}) with
\be
a(h_1,h_2,h_\PPP)=-\frac{i}{2}\eta
\q \bar a(h_1,h_2,h_\PPP)= -\frac{i}{2}{\bar \eta}\,.
\ee

Thus,  the current deformation of massless equations resulting
from the nonlinear HS equations is shown to reproduce the standard local
current interactions up to improvement
$
\PPP_{impr}(w,\PPP)
$
which can be compensated by the field redefinition
\be
\label{2p}
C\to C' = C +\Phi_{1\eta} +\Phi_{2\eta} +\Phi_{1\bar \eta} +\Phi_{2\bar \eta}\,.
\ee

The flip of the $\bar y$ dependence of $h$ in (\ref{F}) to the $y$ dependence in
(\ref{hh}) has an important consequence.
Since  contribution (\ref{F1}) to the spin-one current contains a factor of
$\bar\eta$ while (\ref{hh}) contains a factor of $\eta$ they combine into the
coupling constant $f$ (\ref{nn}) while the phase of $\eta$ cancels out.
Analogously to the spin-one example of Section \ref{deform},
taking into account that left and right components of 0-forms
contribute to the gauge field sector with the additional factors
 $\eta k$ and $\bar \eta \bar k$  as in (\ref{CON1k}) we conclude that
 the physical current  contributing to the gauge field sector is proportional
 to $\eta\bar\eta$. Its explicit form is presented in \cite{Prep}.

The conclusion  that current interactions resulting from
the nonlinear HS equations are independent of the phase of $\eta$
differs drastically from that of \cite{Boulanger:2015ova,Skvortsov:2015lja}
where it was claimed that the result (though divergent) depends on the phase
of $\eta$  and even  vanishes or changes a sign for certain phases.

An important question is which  field redefinitions (\ref{red})
can be treated as local and which cannot. The transformation induced by
$\Phi_2$ (\ref{phi1}) is genuinely local containing a finite number of
derivatives. That induced by $\Phi_1$ is not local in the standard
sense, containing an infinite tale of derivatives.
We postpone the discussion of this issue till the comprehensive
paper \cite{loc} showing in particular that the nonlinear field
redefinition described by $\Phi_1$ is fixed uniquely by the locality condition
at the higher orders.

\section{Holographic interpretation}
\label{hol}
Naively, holographic interpretation of the obtained results may look problematic
since the current interactions resulting from the nonlinear HS equations turn out
to be independent of the phase  of $\eta$. One may wonder how the variety
of models conjectured to be dual to HS theory with different $\eta$
\cite{Aharony:2011jz,Giombi:2011kc} can be generated
by the same lowest-order vertex. This seeming paradox gets
a natural resolution by noticing that different boundary models result from
different boundary conditions in the same bulk model.
Within the approach of \cite{Vasiliev:2012vf,Vasiliev:2015mka} this works as follows.

\subsection{Boundary conditions}
\label{bc}
Due to  dependence
 on the Klein operators $k$ and $\bar k$ the field pattern in the $AdS_4$ HS theory
is doubled. Referring for more detail to sections 8-10 of
\cite{Vasiliev:2012vf} the asymptotic behaviour of the Weyl forms is
\be \label{ct} C^{j\,1-j}(y,\bar y|\xx,{\mathbf z}) = {\mathbf z}
\exp  (y_\ga \bar y^\ga )T^{j\,1-j}(w, \bar w|\xx,{\mathbf z})\,,
\ee
where $C^{j\,1-j}$ is defined in (\ref{cij}),
${\mathbf z}$ is the Poincar\'e coordinate, \be \label{w}
w^\ga = {\mathbf z}^{1/2} y^\ga \q \bar w^\ga = {\mathbf z}^{1/2}
\bar y^\ga\, \ee
and the 0-forms $T^{j\,1-j}$ are associated with
the boundary currents.

$AdS/CFT$ correspondence assumes  certain boundary conditions
at infinity $\zz\to 0$. In terms of $AdS_4$ HS Weyl forms they relate
${ T^{j\,1-j}}(w, 0 | \xx,0)$ to ${ T^{1-j\,j}}(0,{iw} | \xx,0)$.
Indeed, in Eqs.~(8.28), (8.29) of \cite{Vasiliev:2012vf}, the  combinations of currents
\be
\label{calt}
{\bar \eta T^{j\,1-j}}(w, 0 | \xx, 0)-
{ \eta T^{1-j\,j}}( 0,{i w} | \xx,0)\,
\ee
were shown  to contribute the {\it r.h.s.s}
of the equations for HS connections in the boundary limit $\mathbf z\to 0$. As a result, the
contribution of HS connections at the boundary cannot be neglected except
for the boundary conditions
\be
\label{beta}
{\bar \eta T_+^{j\,1-j}}(w, \bar w | \xx,0)- { \eta T_-^{1-j\,j}}( -i \bar w,{i w} |
\xx,0)=0\,,
\ee
where $T_+$ and $T_-$ are the positive and negative helicity
parts of $T(y,\bar y|x)$, respectively.\footnote{I acknowledge with gratitude
a useful discussion of this option with Slava Didenko and Zgenya Skvortsov during
the KITP program New methods in nonperturbative quantum field theory (Jan 6 - Apr 11 2014).}
Note that the relative sign of the factors of $i$ in the arguments of the second term
of (\ref{beta}) is determined by the condition that the fields $T_+^{j\,1-j}(w, \bar w | \xx,0)$
obey the unfolded equations (8.18) of \cite{Vasiliev:2012vf} in the coordinates $\xx$
\be
\Big [d_\xx -i dx^{\ga\gb}\frac{\p^2}{\p w^{+\ga} \p  w^{-\gb}} \Big ] T^{i\,1-j}(w^+, w^- |\xx,\zz)=0.
\ee

As stressed in \cite{Vasiliev:2012vf}, the decoupling of boundary HS
connections is necessary for the proper interpretation of the
boundary model in terms of CFT. If boundary HS connections do couple
to the HS currents, being conformally invariant, the boundary model
is not a CFT in the standard sense lacking the gauge
invariant stress tensor. Generally, one can use arbitrary boundary
conditions reproducing different conformally invariant boundary
field theories. However most of them,
except for those associated with  boundary conditions
(\ref{beta}), are not usual CFT. Such models should
correspond to conformal models considered in \cite{Giombi:2013yva}  where  the
effects of boundary HS gauge fields  in the $AdS_4/CFT_3$ HS duality
were explored.

For the $A$-model ($\eta=1$) and  $B$-model ($\eta=i$) that respect parity,
conditions (\ref{beta}) are equivalent to
\be
\label{dir}
{\mathcal A}^{jj}(w^+, w^- |  \xx) :=
{ T^{j\,1-j}}(w^+,  w^- |  \xx,0)-
{T^{1-j\,j}}(-i  w^-,i  w^+ |  \xx,0)=0\,,
\ee
\be
\label{neum}
{\mathcal B}^{jj}(w^+, w^- | \xx ): =
{ T^{j\,1-j}}(w^+,  w^- | \xx,0)+
{ T^{1-j\,j}}( -i  w^-, i w^+ | \xx,0)=0\,
\ee
called in \cite{Vasiliev:2012vf} \Di\, and \Ne\, conditions, respectively.
Being based on the parity automorphism of the
nonlinear HS theory, the \Di\, condition at $\eta =\bar \eta $ and
\Ne\, condition at $\eta =-\bar \eta $ can be imposed in all orders of the perturbative
expansion. This conclusion  matches the well-known fact
\cite{Klebanov:2002ja,Leigh:2003gk,Sezgin:2003pt} that the $A$-model with
\Di\, boundary conditions and $B$-model with \Ne\, conditions correspond to
free conformal bosons and fermions on the boundary (see also \cite{Giombi:2012ms}
and references therein).
For any other boundary conditions and/or phases of $\eta$ the boundary dual theory
is not free. Hence  the full HS theory containing all boundary models as specific
reductions is not equivalent to a free boundary theory (cf \cite{Maldacena:2011jn}).

 For general $\eta$, the  HS theory is not $P$-symmetric.
 As a result, conditions (\ref{beta}), which distinguish between
positive and negative helicities, break $P$-symmetry.
It is convenient to use following  \cite{Vasiliev:2012vf} the mirror extension of the Poincar\'e coordinate $\zz$ to
negative and, more generally, complex values allowing to extend conditions (\ref{beta})
to
\be
\label{betaz}
{\bar \eta T_+^{j\,1-j}}(w, \bar w | \xx,\zz)- { \eta T_-^{1-j\,j}}( -i \bar w,{i w} |
\xx,-\zz)=0\,.
\ee
These conditions relate self-dual fields in the original space with  anti-self-dual fields
in the mirror space and vice versa in a way manifestly dependent on the phase of $\eta$.
Note that the form of the
extension of  conditions (\ref{beta}) with arbitrary phase of
$\eta$ to higher orders is not obvious and should be derived perturbatively.

Reduction of the universal HS vertex found in this paper to different
subsectors associated with boundary conditions (\ref{beta})
reproduces the conformal structures identified by Maldacena
and Zhiboedov in \cite{Maldacena:2011jn}. Indeed,
after imposing  the boundary conditions (\ref{beta}) the remaining
real boundary fields are
\be
\label{j}
j^j (w^+,w^-|\xx): = \half \left (
\bar \eta T_+^{j\,1-j}(w^+, w^- | \xx,0)+ { \eta T_-^{1-j\,j}}(- i w^-,{i w^+} |
\xx,0)\right )= \bar \eta T_+^{j\,1-j}(w^+, w^- | \xx,0)\,.
\ee

The fact that the bulk HS vertex is independent of the phase $\varphi$
of $\eta$ implies that the boundary vertex  (to be represented by the Lagrangian 4-form  in terms
of \cite{Vasiliev:2015mka}) has  the structure
\be
V = \sum_{i,j=1,2} ( a_{ij} T^{i\,1-i}_+ T^{j\,1-j}_+ +b_{ij} T^{i\,1-i}_- T^{j\,1-j}_-
+e_{ij} T^{i\,1-i}_- T^{j\,1-j}_+)\,,
\ee
where $a_{ij}$, $b_{ij}$ and $e_{ij}$ are some $\varphi$-independent coefficients built
from $\zz$-odd components of derivatives of  the boundary HS connections and background
fields (indices  are implicit).
In terms of real $\varphi$-independent currents (\ref{j}) $V$ reads
\be
\label{V}
V = \f{1}{\eta \bar \eta}\sum_{i,j=1,2} \big( \exp {2i\varphi}\,
a_{ij} j^{i\,1-i}_+ j^{j\,1-j}_+ + \exp {-2i\varphi}\,b_{ij} j^{i\,1-i}_- j^{j\,1-j}_-
+e_{ij} j^{i\,1-i}_- j^{j\,1-j}_+\big)\,.
\ee

Since the dependence on $\varphi$ in (\ref{V}) is manifest we can identify
the parity-even boson and fermion vertices as those associated with
$\varphi=0$ and $\varphi = \pi/2$, respectively. This gives
\be
\label{Vb}
V_{b} =\f{1}{\eta \bar \eta}\sum_{i,j=1,2} (
a_{ij} j^{i\,1-i}_+ j^{j\,1-j}_+ + \,b_{ij} j^{i\,1-i}_- j^{j\,1-j}_-
+e_{ij} j^{i\,1-i}_- j^{j\,1-j}_+)\,,
\ee
\be
\label{Vf}
V_{f} =\f{1}{\eta \bar \eta}\sum_{i,j=1,2} (
-a_{ij} j^{i\,1-i}_+ j^{j\,1-j}_+ - \,b_{ij} j^{i\,1-i}_- j^{j\,1-j}_-
+e_{ij} j^{i\,1-i}_- j^{j\,1-j}_+)\,.
\ee
 Since parity transformation exchanges the left and right sectors,
\ie positive and negative helicities, the remaining parity-odd vertex is
\be
\label{Vo}
V_{o} =\f{i}{\eta \bar \eta}\sum_{i,j=1,2} (
a_{ij} j^{i\,1-i}_+ j^{j\,1-j}_+ - \,b_{ij} j^{i\,1-i}_- j^{j\,1-j}_- )\,.
\ee
This results in  the following formula
\be
\label{zm}
V =   cos^2 (\varphi) V_b +
sin^2 (\varphi) V_f +\half sin(2\varphi)V_{o}\,,
\ee
which precisely matches the form of the deformation of the
HS current algebra found in \cite{Maldacena:2012sf} with $sin^2 (\varphi)=
(1+\tilde \lambda^2)^{-1}$ being in agreement with the $AdS/CFT$ expectation
and the conjecture of Giombi and Yin \cite{G}.

To summarize, the proper dependence on the phase parameter in the
holographic duals of the  $AdS_4$ HS
theory is reproduced by the phase-independent vertex of the bulk  HS
theory derived in this paper via imposing appropriate phase-dependent boundary
conditions.

\subsection{Boundary limit}
\label{limit}
The important question is what is the role of the nonlinear field redefinition
(\ref{2p}) from the boundary point of view? In particular, is it local or nonlocal
in the boundary limit? Naively, formula (\ref{w}) suggests that higher-derivative
contributions resulting from infinite differentiations of the spinorial
variables $y$ and $\bar y$ in the arguments of $T^{j\,1-j}(w,
\bar w|\xx,{\mathbf z})$ are suppressed by additional powers of ${\mathbf z}$
which disappear at ${\mathbf z}\to 0$ so that only some local parts can survive.
However, the story is less trivial.

The important issue here is what happens with the ${\mathbf z}$-independent
factor in (\ref{ct})
\be
F:=\exp  (y_\ga \bar y^\ga )\,.
\ee
As shown in \cite{Vasiliev:2012vf}, $F$ is a Fock projector
\be
F*F = F\,.
\ee
A potential difficulty is that
\be
\tilde F := k *F*k = \exp  (-y_\ga \bar y^\ga )\,,
\ee
which is also a Fock projector
\be
\tilde F*\tilde F = \tilde F\,,
\ee
has divergent product with $F$
\be
F*\tilde F =\tilde F * F =\infty\,.
\ee
This has a consequence that  the homotopy integral
\be
\int d\tau_1 d\tau_2 \rho(\tau)
\exp \tau_1 y_\ga \bar y^\ga * \exp \tau_2 y_\ga \bar y^\ga
\ee
may diverge at $\tau_1, \tau_2 \to -1$ that can lead to
divergencies in the boundary limit. (For more detail see
\cite{Vasiliev:2012vf,Vasiliev:2015mka}.) However, as shown recently in
\cite{Didenko:2017lsn}, for the solution found in this paper
the divergency of this type is absent.

\section{Discussion}
\label{disc}
Confining ourselves to the 0-form sector, we have shown how
nonlinear HS equations in $AdS_4$ reproduce standard local current interactions upon
a proper field redefinition.  This prescription has to be
compared with other suggestions proposed in recent papers
\cite{Boulanger:2015ova,Bekaert:2015tva,Skvortsov:2015lja}.

Conceptually, our conclusions are just opposite to those of
the authors of \cite{Boulanger:2015ova,Skvortsov:2015lja} who
arrived at the conclusion that there are difficulties with the nonlinear
HS equations exhibiting a ``naked singularity" in the analysis of locality,
leading the authors of \cite{Boulanger:2015ova} to counterintuitive  conclusions
like, e.g., the cancellation of the current contributions at $\eta = \exp {\f{i\pi}{4}}$.
On the other hand, our conclusion is that nonlinear HS equations properly reproduce
usual current interactions of massless fields with the coupling constant
independent of the phase of $\eta$.
(Strictly speaking we have  shown this directly only  for
the spin-one currents. However, because interactions of various spins are
controlled by the HS symmetry, the same is true for all other spins as
 shown explicitly in \cite{Prep}.)

The mechanism underlying  cancelation of the phase of
$\eta$ is tricky. Naively, that the arguments of the vierbein $h$ in (\ref{CC})
depend on $\bar y$   suggests that
this part of the deformation should contribute to the right sector of $\bar y$. However
this is not the case since the first field redefinition (\ref{phi1})
leads to an expression with   $h$ independent of both
 $\bar y$ and $y$ while the second one (\ref{phi2}) results in deformation (\ref{hh}) with $h$
depending on $ y$. Hence the
nontrivial part of  deformation (\ref{CC}) lives in the left sector.
This  flip of chirality upon the field redefinition effectively replaces one of the factors
of $\eta$ by $\bar \eta$ and vice versa.

Nevertheless, as shown in Section \ref{hol} along the lines of \cite{Vasiliev:2012vf},
the conclusion  that the HS current interactions
depend on $\eta \bar \eta$  is in agreement with
 various  conjectures on holographic duality of the HS theory with different $\eta$
\cite{Aharony:2011jz,Giombi:2011kc} as well as the Maldacena-Zhiboedov cubic
correlator \cite{Maldacena:2012sf} since the proper dependence on $\eta$ of
the boundary models results from
$\eta$-dependent boundary conditions in the same bulk model.

Since HS theory is based on HS symmetries it is crucially important
to use a gauge invariant setup for the analysis of holography. This concerns
not only the usual gauge symmetries associated with Fronsdal fields but also
the Stueckelberg-like symmetries that control the dependence on the spinorial
variables $Z^A$ (see Section \ref{Nonlinear Higher-Spin Equations}).
To the best of our knowledge, the only available proposal of this type to HS holography
is that of \cite{Vasiliev:2015mka} which is currently under study \cite{dimiv}.
{\it A priori}, it could happen that the field redefinition bringing the current interactions to the
canonical form can result from an allowed gauge transformation in the full space
of $Z,Y$ spinorial variables. In that case it simply will not contribute to the
final result.

For simplicity, in this paper we  focus on the sector of 0-forms which
directly reproduces currents of lower spins $s_\PPP\leq 1$ and implicitly
(\ie via descendants) currents of all higher spins. The fact that it turns
out to be possible to derive local current interactions in this sector suggests
unambiguously that the same should be true in the 1-form sector
to which primary HS currents contribute.
Details of the analysis of this problem are given in \cite{Prep}. In particular,
results of \cite{Prep} simplify  direct comparison of the relative coefficients in front of contributions
of currents of different spins with those derived recently in \cite{Sleight:2016dba}
for the case of $\eta=1$ from the holographic
principle. The two are anticipated  to coincide because at least for the sectors
associated with free boundary theories the relative coefficients are
determined by the HS symmetry. Recently this coincidence  was
explicitly checked in \cite{Misuna:2017bjb}.

The simple  and elegant representation of the field redefinition
 bringing HS interactions to
the canonical current form in terms  of an integral
over  certain simplex  suggests that there should exist a deeper way to
understand HS perturbations in terms of certain geometry in the space of the homotopy
parameters. Further understanding of this issue may be very interesting and suggestive
from both HS and geometric perspectives.

We believe that the results of this paper properly illustrate efficiency of the methods of
\cite{Vasiliev:2012vf,Vasiliev:2015mka} in application to HS holographic duality.
It would be extremely interesting to reproduce the generating functional for boundary
correlators directly from the invariant functional of \cite{Vasiliev:2015mka}
which problem is currently under study \cite{dimiv}.

\section*{Acknowledgements}
I am grateful to Olga Gelfond for many useful discussions and comments, Slava Didenko
for the stimulating discussion of the boundary limit, and Simone Giombi
for the communication.
This research was supported by the Russian Science Foundation Grant No 14-42-00047.

\end{document}